\documentclass[10pt,twoside,twocolumn]{IEEEtran}%
\usepackage{latexsym,delarray,multicol}
\usepackage{epsfig}
\usepackage{amsfonts}
\usepackage{amsthm}
\usepackage{amsmath}
\usepackage{amssymb}
\usepackage{revsymb}
\usepackage{graphicx}
\usepackage{url}
\usepackage{datetime}
\newdateformat{specialdate}{\THEYEAR-\twodigit{\THEMONTH}-\twodigit{\THEDAY}}
\date{\specialdate\today}%
\providecommand{\U}[1]{\protect\rule{.1in}{.1in}}
\newcommand{\nc}{\newcommand}
\nc{\nn}{\nonumber} \nc{\ep}{\varepsilon} \nc{\la}{\lambda}
\nc{\wht}{\widehat} \nc{\ov}{\overline} \nc{\ds}{\displaystyle}
\nc{\ts}{\textstyle}
\nc{\kro}{\left(}\nc{\kvo}{\left[}\nc{\fio}{\left\{}
\nc{\krz}{\right)}\nc{\kvz}{\right]}\nc{\fiz}{\right\}}
\newtheorem{theorem}{Theorem}
\newtheorem{lemma}[theorem]{Lemma}

\begin{document}

\title{Polar codes with a stepped boundary}
\author{Ilya Dumer \vspace
{0.05in} \\ February 15, 2017\thanks{I. Dumer is with the College of Engineering, University of
California, Riverside, CA 92521, USA; email: dumer@ee.ucr.edu }}
\date{\today}
\maketitle

{\small \noindent}\textbf{Abstract: }We consider explicit polar constructions
of blocklength $n\rightarrow\infty$ for the two extreme cases of code rates
$R\rightarrow1$ and $R\rightarrow0.$ For code rates $R\rightarrow1,$ we design
codes with complexity order of $n\log n$ in code construction, encoding, and
decoding. These codes achieve the vanishing output bit error rates on the
binary symmetric channels with any transition error probability $p\rightarrow
0$ and perform this task with a substantially smaller redundancy $(1-R)n$ than
do other known high-rate codes, such as BCH codes or Reed-Muller (RM). We then
extend our design to the low-rate codes that achieve the vanishing output
error rates with the same complexity order of $n\log n$ and an asymptotically
optimal code rate $R\rightarrow0$ for the case of $p\rightarrow1/2.\vspace
{-0.01in}$


\section{\textbf{Introduction}}

Below we consider the \textsf{Plotkin }recursive construction $\mathbf{u,u+v}$
that repeatedly combines shorter codes to construct and decode the longer
ones. \ RM codes $\mathcal{R}(r,m)$ represent one Plotkin-type construction
\cite{MWS81} of length $n=2^{m}$ and dimension $k(r,m)=\sum_{0}^{r}\left(
_{\,i}^{m}\right)  $ with parameters $0\leq r\leq m.$ \ Polar codes \cite{ari}
introduce another recursive design. \ Both codes originate from the same
full--space code $\mathcal{R}(m,m)$ and filter it in two different ways.
\ Namely, a code $\mathcal{R}(r,m)$ maximizes the code rate among all codes
that have the same distance $2^{m-r}$ and are generated by the $m$-variate
Boolean monomials. Polar codes use a more intricate optimization. First, the
successive-cancellation decoding (SCD) of \cite{dum1}-\cite{dum4} performs
step-by-step retrieval of information bits \ of code $\mathcal{R}(m,m).$
Analysis of SCD \cite{dum4} shows that it yields both high and low-fidelity
information bits for RM codes. Therefore, removing low-fidelity bits (by
setting them as zeros) gives the better-performing subcodes of RM codes. For
relatively short lengths of 512 or less, this was done in \cite{Dum2001,dum4}.
In particular, it turns out that these subcodes achieve a nearly optimal
(ML)\ performance on these lengths if SCD is combined with list decoding. For
long codes with $m\rightarrow\infty$, the major breakthrough achieved in
\cite{ari} shows that the subcodes of $\mathcal{R}(m,m)$ that keep $Rn$ most
reliable bits are capacity achieving (CA) codes under SCD for any binary
symmetric memoryless channel $U$ and any code rate $R\in(0,1).$ These\textsf{
polar }codes also achieve a polynomial complexity of construction. Namely, for
a channel $U$ with capacity $C,$ polar codes of code rate $R>C-\epsilon$ have
complexity \cite{guru-2015} of order \textsf{poly}$\left(  a\epsilon^{-\mu
}\right)  $ for any $\epsilon>0,$ where $a=a\left(  U\right)  $ and $\mu$ are
some constants.

Below, we extend the above results for the special cases of $R\rightarrow1$
and $R\rightarrow0.$ In both cases, we consider code families that achieve a
vanishing output bit error rate on a binary symmetric channel BSC$(p)$ with a
transition error probability $p$ and capacity $C=1-h(p),$ where $h(p)$ is a
binary entropy. We say that a family of codes with $n\rightarrow\infty$ and
$R\rightarrow1$ is \textsf{strongly optimal} if the fraction $\rho=1-R$ of
redundant (parity-check) bits has the smallest possible order%
\[
1-R\sim h(p)=p\log_{2}({}^{e}\!\!\left/  _{p}\right.  )+O(p^{2})
\]
A family of long codes is called \textsf{weakly optimal }if probability
$p\rightarrow0$ and redundancy $\rho$ have a similar decline rate%
\begin{equation}
\log_{2}(1-R)\sim\log_{2}h(p)\sim\log_{2}p \label{weak}%
\end{equation}
Our main result is as follows.

\begin{theorem}
\label{th:1}For any $p\rightarrow0,$ there exist weakly optimal codes of
length $n\rightarrow\infty$ that have a relative redundancy%
\begin{equation}
\rho\leq p\left(  \log_{2}{}^{1}\!\!\left/  _{p}\right.  \right)  ^{\log
_{2}\log_{2}\text{ }{}^{1}\!\!\left/  _{p}\right.  } \label{main}%
\end{equation}
and achieve a vanishing error probability on a binary symmetric channel
BSC$(p).$ These codes can be constructed, encoded, and decoded with complexity
of order $n\ln n.$
\end{theorem}

Similarly, long codes of rate $R\rightarrow0$ are called strongly optimal if
they achieve a vanishing output error rate on a BSC$(p)$ with $p\rightarrow
1/2$ and have the maximum possible order of code rate $R\sim1-h(p)\sim
(1-2p)^{2}/\ln4.$ We extend Theorem 1 and design strongly optimal codes of
rate $R\rightarrow0$ and complexity $n\ln n.$

For a wide range of error probabilities $p,$ codes of Theorem 1 outperform
known codes of code rate $R\rightarrow1.$ For example, long primitive BCH
codes require redundancy $p\log_{2}n$ to achieve a vanishing output error rate
under the bounded-distance decoding on a BSC$(p)$ if $p=o(\log_{2}n)$
\cite{MWS81}$.$ However, $R\rightarrow0$ if $p\log_{2}n\rightarrow\infty.$ The
recent breakthrough of \cite{shpilka-2016} also shows that high-rate RM codes
$\mathcal{R}(m-2r-1,m)$ can correct the fraction of errors $p\sim\left(
_{\,r}^{m}\right)  /2^{m}$ with polynomial complexity and low redundancy
$\rho\sim\left(  _{2r+1}^{\,\,m}\right)  /2^{m}$ if $r=o(\sqrt{m/\log m}).$
This algorithm is still limited to the rapidly vanishing probabilities $p$
unlike any $p\rightarrow0$ in Theorem 1. Note, however, that Theorem 1
achieves no improvements over BCH codes if probability $p$ has an
exponentially declining order $p\leq2^{-m^{c}}$ for any $c>0$, nor does it
give strongly optimal codes for $R\rightarrow1.$

Sections \ref{sub:2} and \ref{sub:3} provide some background and address the
common properties of RM and polar codes. Sections \ref{sub:4}-\ref{sub:6}
introduce polarized design with a single boundary. We first design the weakly
optimal codes of rates $R\rightarrow1$ and then extend them to the strongly
optimal codes of rate $R\rightarrow0.\vspace{-0.06in}$

\section{\textbf{Recursive design of RM and polar codes}\label{sub:2}}

Consider boolean polynomials $f(x)$ of degree $r$ or less in $m$ binary
variables $x_{1},\ldots,x_{m}$, where $r\leq m$.\ Vectors $x=(x_{1}%
,...,x_{m})$ will mark the positions of our code. Each map $f(x):\mathbb{F}%
_{2}^{m}\rightarrow\mathbb{F}_{2}$ generates a codeword $\mathbf{c=c}(f)$ of
code $\mathcal{R}(r,m).$ \ We also use short notation $\mathbf{x}%
_{i\,|\,j}=(x_{i},...,x_{j})$ for $i\leq j.$ Consider recursive decomposition%
\begin{equation}%
\begin{tabular}
[c]{l}%
$f(x)=f_{0}(\mathbf{x}_{2\,|\,m})+x_{1}f_{1}(\mathbf{x}_{2\,|\,m})=...$\\
$=\sum_{i_{1},...,i_{\ell}}x_{1}^{i_{1}}\cdot...\cdot x_{\ell}^{i_{\ell}%
}\,\,f_{i_{1},...,i_{\ell}}(\mathbf{x}_{\ell+1\,|\,m})$\\
$=...=\sum_{i_{1},...,i_{m}}f_{i_{1},...,i_{m}}\;x_{1}^{i_{1}}\cdot...\cdot
x_{m}^{i_{m}}$%
\end{tabular}
\ \ \ \ \ \label{poly1}%
\end{equation}
The first step decomposes \ polynomial $f(x)$ into polynomials $f_{0}$ and
$f_{1}$ of degrees $\deg f_{0}\leq\min\{r,m-1\}$ and $\deg$ $f_{1}\leq r-1.$
Then the codewords $\mathbf{c}_{0}=\mathbf{c}(f_{0})$ and \ $\mathbf{c}%
_{1}=\mathbf{c}(f_{1})$ belong to the codes $\mathcal{R}(r,m-1)$ and
$\mathcal{R}(r-1,m-1)$ and form the codeword $\mathbf{c=c}_{0}\mathbf{,c}%
_{0}\mathbf{+c}_{1}$ of code $\mathcal{R}(r,m)$. \ Similarly, any subsequent
step $\ell$ decomposes each polynomial with respect to $x_{\ell}^{i_{\ell}}$
as follows%
\[
f_{i_{1},...,i_{\ell-1}}(\mathbf{x}_{\ell\,|\,m})=\sum\nolimits_{i_{\ell}%
=0,1}f_{i_{1},...,i_{\ell}}(\mathbf{x}_{\ell+1\,|\,m})\cdot x_{\ell}^{i_{\ell
}}%
\]
We then say that the $\ell$-level binary \textit{paths} $\xi_{1\,|\,\ell
}=i_{1},...,i_{\ell}$ decompose the original polynomial $f(x)$ into sums of
monomials $x_{1}^{i_{1}}\cdot...\cdot x_{\ell}^{i_{\ell}}f_{i_{1},...,i_{\ell
}}(\mathbf{x}_{\ell+1\,|\,m}).$ \ Finally, \ full \textsf{paths} $\xi
=i_{1},...,i_{m}$ of step $m$ \ define monomials $x^{\xi}\equiv x_{1}^{i_{1}%
}\,\cdot...\cdot\,x_{m}^{i_{m}}$ with coefficients $f_{\xi}=f_{i_{1}%
,...,i_{m}}=0,1.$ Note that each monomial $x^{\xi}$ gives a codeword
$\mathbf{c}(x^{\xi})$ of \ weight $2^{m-w(\xi)},$ where $w(\xi)$ is the
Hamming weight of the string $\xi.$ RM codes $\mathcal{R}(r,m)$ include only
$k(r,m)$ paths of weight $w(\xi)\leq r.$

In Fig. 1 we use this representation for the full code $\mathcal{R}(4,4)$.
Each decomposition step $\ell=1,...,4$ is marked by the splitting monomial
$x_{\ell}^{i_{\ell}}.$ For example, path $\xi=0110$ gives the coefficient
$f_{0110}$ \ associated with the monomial $x^{\xi}\equiv x_{2}x_{3}$.

Fig. 2 depicts code $\mathcal{R}(2,5).$ Here we only include all paths
$\mathbf{\mathbf{\mathbf{\xi}}}$ of weight $w(\xi)\leq2.$ Note that any two
paths $\xi_{1\,|\,\ell}$ entering some node have the same weight $w$ and
generate the same code $\mathcal{R}(r-w,m-\ell)$ on their extensions. For
example, path $\xi=01100$ proceeds from $\mathcal{R}(2,5)$ to the single bit
$\mathcal{R}(0,0)$ via codes $\mathcal{R}(2,4),$ $\mathcal{R}(1,3),$
$\mathcal{R}(0,2),$ and $\mathcal{R}(0,1).$

\begin{figure}[ptb]
\vspace{-0.1in}
\includegraphics[width=0.48\textwidth]{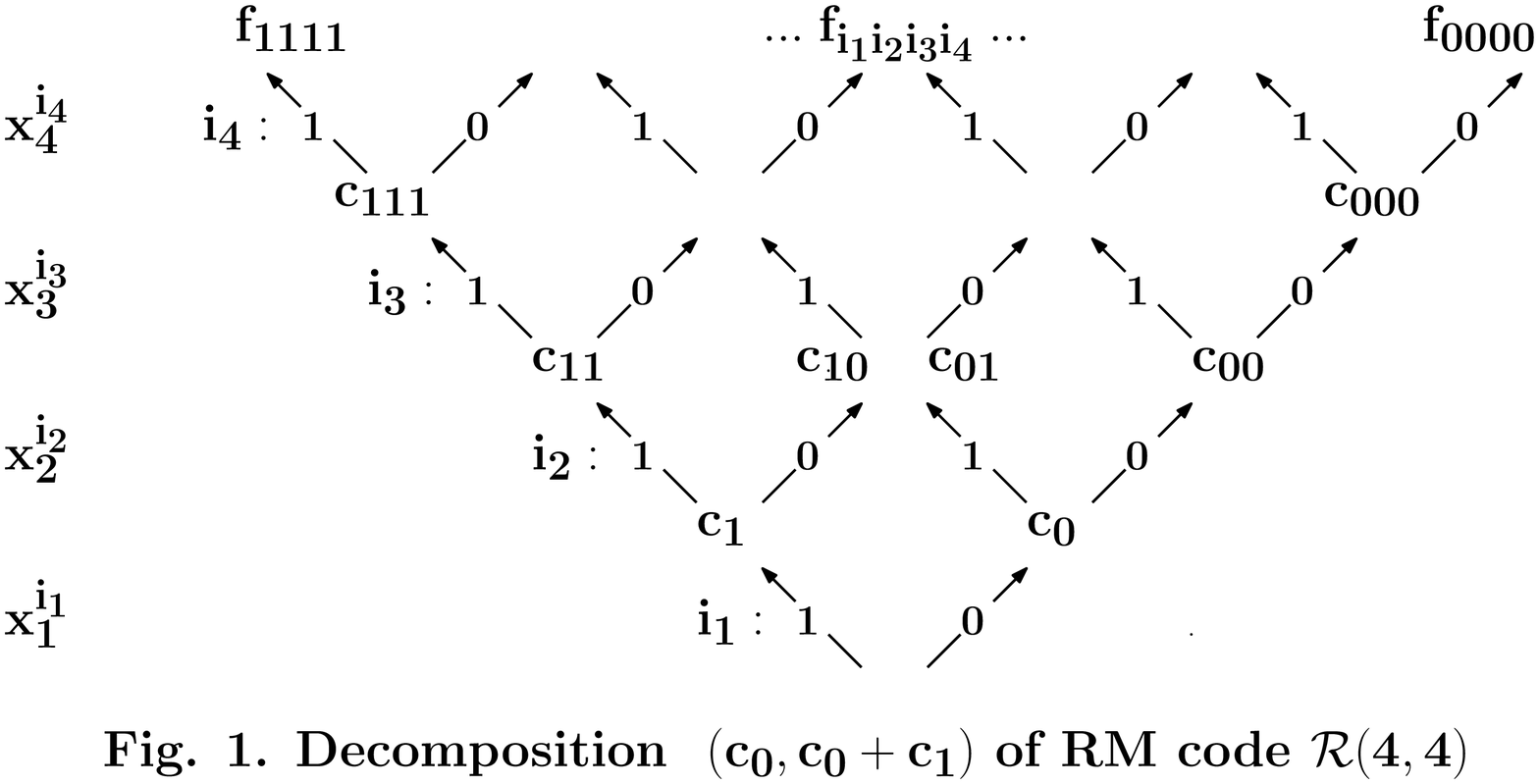}\end{figure}\begin{figure}[ptb]
\includegraphics[width=0.41\textwidth]{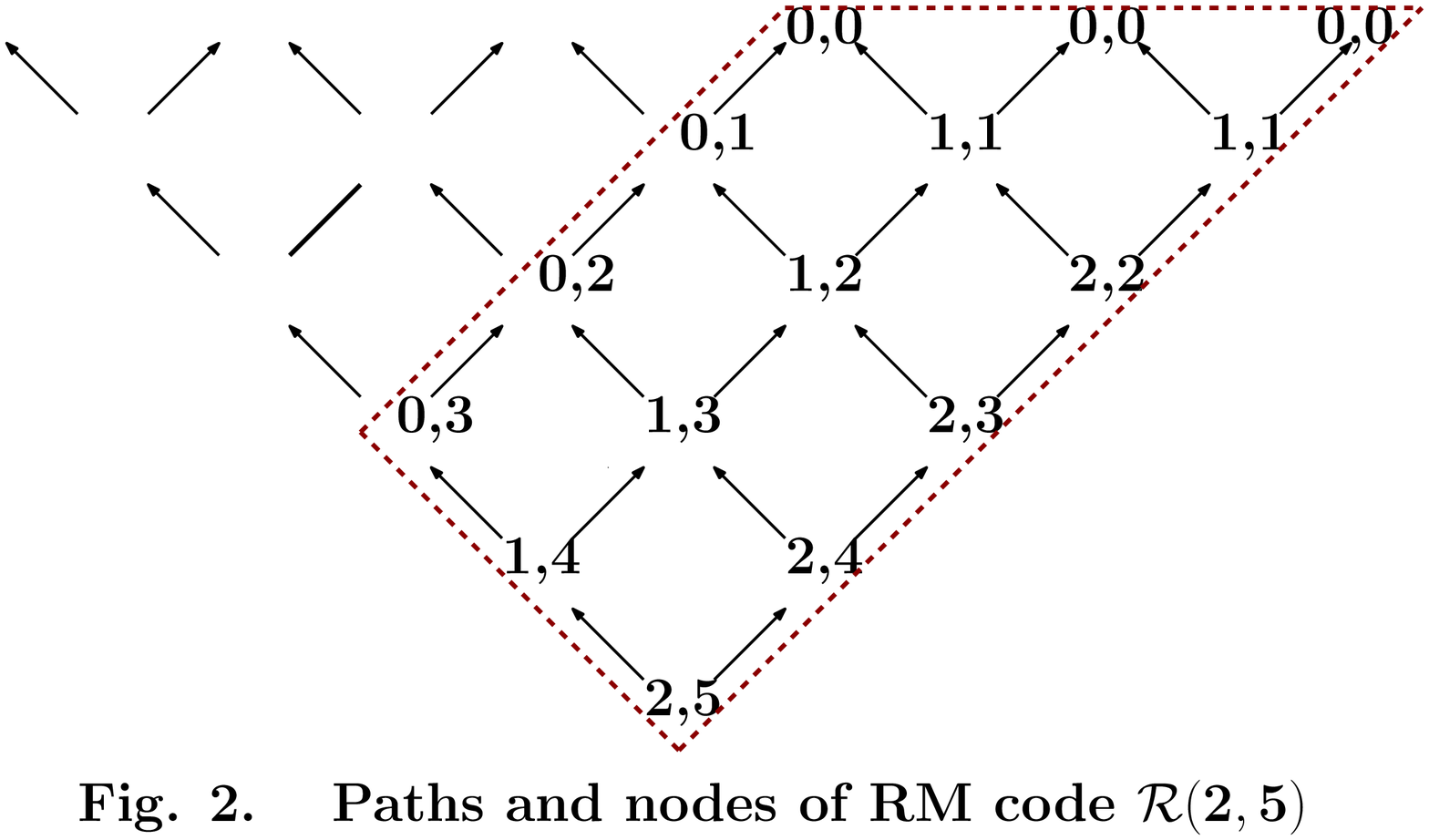}\end{figure}

This design can be reformulated using a $2\times2$ matrix%
\[
G=\left[
\begin{array}
[c]{cc}%
1 & 1\\
0 & 1
\end{array}
\right]
\]
Then code $\mathcal{R}(m,m)$ is generated by the Kronecker product
$G(m,m)=G^{\otimes m}.$ Each row of $G^{\otimes m}$ is the map of the monomial
$x^{\xi}$ for some path $\xi.$ Similarly, matrix $G(r,m)$ is the map of all
monomials $x^{\xi}$ with paths $\xi=i_{1},...,i_{m}$ of weight $w(\xi)\leq r$.\

Now consider a single path $\xi$ that ends with an information bit
$f_{i_{1},...,i_{m}}=1$. Encoding proceeds in the reverse order $\ell
=m,...,1.$ We begin with a single bit codeword $\mathbf{c(\mathbf{\xi}%
}_{m+1\mid m})=1.$ In each step $\ell,$ we use recursion and obtain the
codeword%
\begin{equation}
\mathbf{c}\left(  \mathbf{\mathbf{\xi}}_{\ell\,|\,m}\right)  =\left\{
\begin{tabular}
[c]{ll}%
$\mathbf{c(\mathbf{\xi}}_{\ell+1\,|\,m}),\mathbf{c(\mathbf{\xi}}%
_{\ell+1\,|\,m})$ & $\;$if $i_{\ell}=0\smallskip$\\
$\mathbf{0,c(\mathbf{\xi}}_{\ell+1\,|\,m})$ & $\;$if $i_{\ell}=1$%
\end{tabular}
\ \ \ \ \ \ \right.  \label{code1}%
\end{equation}
of length $2^{m-\ell+1}.$ Thus, any path $\xi$ is encoded in the vector
$\mathbf{c}=\mathbf{c}(\xi)$ of length $n.$ Also, $\mathbf{c}(\xi)=0$ if
$f_{i_{1},...,i_{m}}=0.$

Now consider a subset of $N$ paths $T.\mathbf{\ }$ Then we encode $N$
information bits via their paths and obtain codewords $\mathbf{c}(T)=\sum
_{\xi\in T}\mathbf{c\mathbf{(\xi).}}$ These codewords form a linear code
$C(m,T).$ Here at any level $\ell,$ encoding adds two codewords of level
$\ell+1$ entering any node $\xi_{\ell\,|\,m}.$ Thus, encoding (\ref{code1})
performs $2^{m-\ell}$ operations on each of $2^{\ell}$ nodes $\xi_{\ell
\,|\,m}$ and has the overall complexity of $n\log_{2}n$ over all levels $\ell$.

\begin{lemma}
\label{lm:0}Code $C(m,T)$ has length $2^{m}$, dimension $|T|$ and distance
$2^{m-r},$ where $r=\max\{w(\xi),\xi\in T\}$ is the weight of the heaviest
path in $T.$ Code $\mathcal{R}(r,m)$ has the maximum code rate $R$ among all
codes $C(m,T)$ of the distance $2^{m-r}.$
\end{lemma}

\noindent\textit{Proof. }Let weight $r$ be achieved on some path $\psi\in T.$
\ Then code $C(m,T)$ is generated by monomials\ $x^{\xi}$ of degree $r$ or
less. Thus, $C(m,T)\subseteq\mathcal{R}(r,m).$ The monomial $x^{\psi}$ has
degree $r$ and gives the minimum weight $2^{m-r}\mathbf{.}$\hfill
${\dimen0=1.5ex\advance\dimen0by-0.8pt\relax\mathop{\mkern0.5\thinmuskip
\vbox{\hrule width 1.5ex
\hbox to 1.5ex{\vrule
\hfill
\vrule height\dimen0 width 0pt
\vrule}\hrule width 1.5ex}\mkern0.5\thinmuskip}}\vspace{-0.06in}$

\section{\textbf{Recursive decoding algorithms\label{sub:3}}}

Below, we use a map $x\rightarrow(-1)^{x}$ for any $x=0,1$ and consider a
discrete memoryless channel (DMC) $W$ with inputs $\pm1.$ Vector $\mathbf{ab}$
will denote the component-wise product of vectors $\mathbf{a},$ $\mathbf{b}$
and $\mathbf{c}=\left(  \mathbf{u,uv}\right)  $ will denote the codewords
$\mathbf{c}$ of a code $\mathcal{R}(r,m)$ with symbols $\pm1\mathbf{.}$ In
particular, $\mathbf{1}^{n}$ now represents a former all-zero codeword$.$\ For
any codeword $\mathbf{c},$ let $\mathbf{y}_{0}$, $\mathbf{y}_{1}$ be the two
output halves corrupted by noise. We use double index $i,j$ for any position
$j=1,...,n/2$ in a half $i=0,1$. \ Define the posterior probability (PP)
$q_{i,j}=\Pr\{c_{i,j}=1\,\,|\,\,y_{i,j}\}$ that $1$ is sent in position $i,j.$
We will often replace $q_{i,j}$ with two related quantities, which we call
\ \textquotedblleft the offsets" $g_{i,j}$ and the likelihoods $h_{i,j}:$%
\begin{equation}
g_{i,j}=2q_{i,j}-1,\;h_{i,j}=q_{i,j}/\left(  1-q_{i,j}\right)  \label{lik}%
\end{equation}
Thus, we will use vectors $\mathbf{q}=(q_{i,j}),$\ \ $\mathbf{g}=(g_{i,j})$
and $\mathbf{h}=(h_{i,j}).$ For example, let $W$ be a binary symmetric channel
BSC$(p),$ where $p=(1-\epsilon)/2.$ Then any output $y=\pm1$ gives quantities
$g(y)=\epsilon y$ and $h(y)=(1+\epsilon y)/(1-\epsilon y)$.

The following recursive algorithm $\Psi_{r}^{m}(\mathbf{q})$ of \cite{dum1},
\cite{Dum2001} performs SCD of information bits in codes $\mathcal{R}(r,m)$ or
their subcodes $C(m,T).$ Here we relegate decoding of vector $\mathbf{q}$ to
two vectors $\mathbf{q}^{(1)}$ and $\mathbf{q}^{(0)}$ of length $n/2.$ Vector
$\mathbf{q}^{(1)}$ consists of PP $q_{j}^{(1)}\equiv\Pr\{v_{j}%
=1\,\,|\,\,q_{0,j},\,q_{1,j}\}$ of symbols $v_{j}$ in construction $\left(
\mathbf{u,uv}\right)  .$ Simple recalculations \cite{dum1} show that the
offsets $g_{j}^{(1)}$ of symbols $v_{j}$ can be expressed as the products of
two offsets $g_{0,j}g_{1,j}.$ Thus, we obtain vectors $\mathbf{g^{(1)}}$ and
$\mathbf{q}^{(1)}$ with symbols%
\begin{equation}
g_{j}^{(1)}=g_{0,j}g_{1,j},\;\;q_{j}^{(1)}=(1+g_{j}^{(1)})/2. \label{1}%
\end{equation}
We may now apply some decoding algorithm $\Psi_{\,r-1}^{m-1}$ to the vector
$\mathbf{q}^{(1)}$ and obtain a vector $\widetilde{\mathbf{v}}\in
\mathcal{R}(r-1,m-1)$ of length $n/2.$ Now we have two corrupted versions
$\mathbf{y}_{0}$ and $\mathbf{y}_{1}\widetilde{\mathbf{v}}$ of vector
$\mathbf{u}.$ We can then derive PP $q_{j}^{(0)}=\Pr\{u_{j}%
=1\,\,|\,\,\,q_{0,j}\,,\,q_{1,j},\widetilde{v}_{j}\}$ of symbols $u_{j}$ in
the $\left(  \mathbf{u,uv}\right)  $ construction. Indeed, any symbol $u_{j}$
has likelihoods $h_{0,j}$ and $\left(  h_{1,j}\right)  ^{\widetilde{v}_{j}}$
in the left and right halves, respectively. Then we combine the two
likelihoods into their product:
\begin{equation}
h_{j}^{(0)}=h_{0,j}\left(  h_{1,j}\right)  ^{\widetilde{v}_{j}},\;\;q_{j}%
^{(0)}=h_{j}^{(0)}/(1+h_{j}^{(0)}) \label{2}%
\end{equation}
Then we can apply some decoding $\Psi_{\,r}^{m-1}$ to vector $\mathbf{q}%
^{(0)}$ and obtain $\widetilde{\mathbf{u}}$ $\mathbf{\in}$ $\mathcal{R}%
(r,m-1).$

Decomposition (\ref{1}), (\ref{2}) forms level $\ell=1$ of \ SCD, which can
also be continued for vectors $\mathbf{q}^{(1)}$ and $\mathbf{q}^{(0)}$ on the
codes $\mathcal{R}(r-1,m-1)$ and $\mathcal{R}(r,m-1).$ Then levels
$\ell=2,...,m$ are processed similarly, moving decoding along the paths of
Fig. 1 or Fig. 2. Any incomplete path $\mathbf{\mathbf{\xi}}_{1\,|\,\ell}$
begins with its $\mathbf{v}$-extension $(\mathbf{\mathbf{\xi}}_{1\,|\,\ell
},1).$ Upon decoding, this path delivers its output $\widetilde{\mathbf{v}}$
to the $\mathbf{u}$-path $(\mathbf{\mathbf{\xi}}_{1\,|\,\ell},0)$. Thus, all
paths are ordered lexicographically. Finally, the last step gives the
likelihood $q_{\mathbf{\mathbf{\xi}}}=\Pr\{f_{\mathbf{\mathbf{\xi}}%
}=0\,\,|\,\,\mathbf{y}_{0},\mathbf{y}_{1}\}$ of one information bit
$f_{\mathbf{\mathbf{\xi}}}$ on the path $\mathbf{\mathbf{\xi.}}$ We then
choose the more reliable bit $f_{\mathbf{\mathbf{\xi}}}.$ It is easy to verify
\cite{dum1} that $m$ decomposition steps give complexity $2n\log_{2}n.$

Any subcode $C(m,T)$ is decoded similarly and assumes that all paths
$\mathbf{\mathbf{\xi\notin}}T$ are frozen and give information bits
$f_{\mathbf{\mathbf{\xi}}}\equiv0.$ \ Let all $N$ paths in $T$ be ordered
lexicographically as $\mathbf{\mathbf{\xi}}^{(1)},...,\mathbf{\mathbf{\xi}%
}^{(N)}$. Then we have \smallskip\smallskip

$\frame{$%
\begin{array}
[c]{l}%
\text{Algorithm }\Psi(m,T)\text{ for code }C(m,T).\smallskip\\
\text{Given: a vector }\mathbf{q}=(q_{i,j})\text{ of PP.}\smallskip\\
\text{Take }s=1,...,N\text{ and }\ell=1,...,m.\smallskip\\
\text{For path }\mathbf{\mathbf{\xi}}^{(s)}=i_{1}^{(s)},...,i_{m}^{(s)}\text{
in step }\ell\text{ do:}\smallskip\\%
\begin{array}
[c]{l}%
\text{Apply recalculations}\;\text{(\ref{1}) if }i_{\ell}^{(s)}=1\smallskip
\smallskip\\
\text{Apply recalculations}\;\text{(\ref{2}) if }i_{\ell}^{(s)}=0.\smallskip\\
\text{Output the bit }f_{\mathbf{\mathbf{\xi}}^{(s)}}\text{ for }\ell=m.
\end{array}
\end{array}
$}\vspace{-0.06in}$

\section{\textbf{Path ordering in SC decoding }\label{sub:4}}

Let a binary code $C(m,T)$ be used over a symmetric DMC $W.$ We now consider a
code $C_{\mathbf{\mathbf{\xi}}}$ defined by a single path $\xi=(i_{1}%
,...,i_{m})$ and estimate its decoding error probability
$P_{\mathbf{\mathbf{\xi}}}.$ Let a codeword $\mathbf{1}^{n}$ be transmitted
over this path. We now may assume that other paths give outputs $\widetilde
{v}_{j}=1$ in recursive recalculations (\ref{lik})-(\ref{2}). Then we
re-arrange (\ref{lik})-(\ref{2}) as follows%
\begin{align}
g_{j}^{(1)}  &  =g_{0,j}g_{1,j},\quad g_{j}^{(0)}=(g_{0,j}+g_{1,j}%
)/(1+g_{0,j}g_{1,j})\smallskip\label{m21}\\
h_{j}^{(0)}  &  =h_{0,j}h_{1,j},\quad h_{j}^{(1)}=(1+h_{0,j}h_{1,j}%
)/(h_{0,j}+h_{1,j})\smallskip\label{m2}%
\end{align}
\ From now on, we may consider recalculations (\ref{m21}) and (\ref{m2}) as
the sequences of channel transformations applied to the original random
variables (rv) $g_{i,j}$ or $h_{i,j}.$ In the end, we obtain a new memoryless
channel $W_{\xi}:$ $X\rightarrow Y_{\mathbf{\mathbf{\xi}}}$ that outputs a
single rv $h(\xi)$ after $m$ steps. For any parameter $\lambda>0,$ we also
consider rv $h^{\lambda}(\xi)$ and its expectation $\mathbb{E}h^{-\lambda}%
(\xi)$. Then the Chernoff upper bound gives%
\[
P_{\mathbf{\mathbf{\xi}}}\equiv\Pr\{h(\xi)<1\}\leq\min_{\lambda>0}%
\mathbb{E}h^{-\lambda}(\xi)=\min_{\lambda>0}\mathbb{E}e^{-\lambda\ln h(\xi)}%
\]
Note that the quantity $\mathbb{E}h^{-1/2}(\xi)$ is identical to the
Bhattacharyya parameter%
\[
Z(W)=\ts\sum_{y\in Y}\sqrt{W(y|0)}\sqrt{W(y|1)}%
\]
defined for a DMC channel $W_{\xi}:$ $X\rightarrow Y_{\mathbf{\mathbf{\xi}}}.$
For example, BSC$(p)$ with $p=(1-g)/2$ gives%
\[
Z(W)=\mathbb{E}h^{-1/2}(\xi)=\ts2\left(  \frac{1+g}{2}\right)  ^{1/2}%
\ts\left(  \frac{1-g}{2}\right)  ^{1/2}=\sqrt{1-g^{2}}%
\]
In a more general setting \cite{korada}, we decompose a binary symmetric DMC
$W_{\mathbf{\mathbf{\xi}}}$ into some number $k$ of binary symmetric channels
BSC$_{\theta_{i}}(p_{i})$ that have transition error probabilities
$p_{i}=(1-g_{i})/2$ and occur with some probability distribution $\{\theta
_{i}\},$ where $\sum_{1}^{k}\theta_{i}=1.$ Then%
\begin{equation}
Z(W_{\xi})=\ts\sum\nolimits_{i}\theta_{i}\sqrt{1-g_{i}^{2}} \label{kor1}%
\end{equation}
Below we use the upper bound $P_{\mathbf{\mathbf{\xi}}}\leq Z(W_{\xi})$
employed by Arikan in \cite{ari}$.$ It is also proved in \cite{ari} that a one
step recursion $(W,W)\rightarrow\left(  W^{(1)},W^{(0)}\right)  $ of
(\ref{m2}) gives parameters $Z(W^{(1)})$ and $Z(W^{(0)})$ such that%
\begin{equation}
1-Z(W^{(1)})\geq\left[  1-Z(W)\right]  ^{2},\quad Z(W^{(0)})=Z^{2}(W)
\label{arik1}%
\end{equation}
Now consider a compound channel $W_{\mathbf{\mathbf{\xi}}}$ as a set of
BSC$_{\theta_{i}}(p_{i}).$ Then we can define the expectation of the offsets
$g_{i}>0:$
\[
\mathcal{G}\left(  W_{\mathbf{\mathbf{\xi}}}\right)  =\sum\nolimits_{1}%
^{k}\theta_{i}g_{i}%
\]
Note that $\sqrt{1-g^{2}}$ is a concave function. Also, $\sqrt{1-g^{2}}%
\geq1-g$ for any $g\in\lbrack0,1].$ Thus, (\ref{kor1}) yields two inequalities%
\begin{equation}
1-\mathcal{G}\left(  W_{\mathbf{\mathbf{\xi}}}\right)  \leq Z(W_{\xi}%
)\leq\sqrt{1-\left[  \mathcal{G}\left(  W_{\xi}\right)  \right]  ^{2}}
\label{dum1}%
\end{equation}
Given a one step recursion $(W,W)\rightarrow\left(  W^{(1)},W^{(0)}\right)  ,$
we can also take two independent identically distributed rv $g_{0,j}$ and
$g_{1,j}$ in (\ref{m21}) and find the expectation of their product
$g_{j}^{(1)}$ for the channel $W^{(1)}.$ Then we have two equalities%
\begin{align}
\mathcal{G}(W^{(1)})  &  =\mathcal{G}^{2}(W),\;\label{dum2}\\
Z(W^{(0)})  &  =Z^{2}(W) \label{dum3}%
\end{align}
Below we replace notation $Z(W_{\xi})$ and $\mathcal{G}(W_{\xi})$ with
$Z(\xi)$ and $\mathcal{G}(\xi).$ Given a path $\xi=(i_{1},...,i_{m}),$ we say
that a path $\eta=(j_{1},...,j_{m})$ is its \textsf{descendant} if $\eta$ is
obtained from $\xi$ by the following replacements in any positions $s$ or
$(s,s+1):~$%
\begin{align}
i_{s}  &  =1\Rightarrow j_{s}=0,\nonumber\\
\left(  i_{s}=1,i_{s+1}=0\right)   &  \Rightarrow\left(  j_{s}=0,j_{s+1}%
=1\right)  \label{des1}%
\end{align}
Let $h(\xi)$ and $h(\eta)$ be the outputs of paths $\xi$ and $\eta$ obtained
by recalculations (\ref{m2})$.$ The following Lemma \ref{lm:tau} uses a
partial order for the paths $\xi$ and $\eta$ with respect to the quantities
$\mathbb{E}h^{-\lambda}(\xi).$ A similar lemma was used in \cite{bd1} for a
slightly different set of recalculations, which approximate recalculations
(\ref{m2}). In \cite{schu} and \cite{till}, this lemma is proved for the
Bhattacharyya parameter $Z(W_{\xi})$ with exact recalculations (\ref{m2}). In
Appendix, we also post a proof of Lemma \ref{lm:tau} for the arbitrary moments
$\mathbb{E}h^{-\lambda}(\xi)$.

\begin{lemma}
\label{lm:tau} Recalculations (\ref{m2}) on some path $\xi$ and its descendant
$\eta$ give the outputs $h(\xi)$ and $h(\eta)$ \ that satisfy inequalities%
\begin{align}
\mathsf{E}h^{-\lambda}(\xi)  &  \geq\mathsf{E}h^{-\lambda}(\eta),\quad
\lambda\in\lbrack0,1],\label{ch2}\\
\mathsf{E}h^{-\lambda}(\xi)  &  \leq\mathsf{E}h^{-\lambda}(\eta),\quad
\lambda\in\lbrack1,\infty). \label{ch3}%
\end{align}
\textbf{Corollary.}\label{lm:cor} Any path $\xi$ and its descendant $\eta$
satisfy inequalities $P(\eta)\leq Z(\eta)\leq Z(\xi).$
\end{lemma}

Below, we say that a path $\xi$ forms \emph{a boundary }for all
descendant\emph{ } paths $\eta$ that satisfy replacements (\ref{des1}%
).$\vspace{-0.06in}$

\section{\textbf{High-rate codes with a stepped boundary}\label{sub:5}}

Below, $\log x\equiv\log_{2}x$. For $i=1,...,s,$ consider a set of $2s$
non-negative ordered integers ${\mathcal{L=}}\left\{  r_{i},\ell_{i}\right\}
$ such that $\,{r}_{i}+\ell_{i}=m_{i}\,\ $and $\sum_{i=1}^{s}m_{i}=m.$ We say
that a path$\smallskip$%
\begin{equation}
\xi\left(  {\mathcal{L}}\right)  =\xi^{(1)},...,\xi^{(s)}=1^{{r}_{1}}%
0^{\ell_{1}},...,1^{{r}_{s}}0^{\ell_{s}} \label{ksi}%
\end{equation}
of length $m$ bounds a path $\eta({\mathcal{L)=}}\eta^{(1)},...,\eta^{(s)}$ if
each section $\eta^{(i)}$ of length $m_{i}$ has weight
\begin{equation}
w(\eta^{(i)})\leq{r}_{i},\,i=1,...,s \label{nu}%
\end{equation}
Thus, each section $\eta^{(i)}$ is located to the right of $\xi^{(i)}$ as seen
in Fig. 3 for a path $\xi\left(  {\mathcal{L}}\right)  =1^{{r}_{1}}0^{\ell
_{1}}1^{{r}_{2}}0^{\ell_{2}}1^{{r}_{3}}0^{\ell_{3}}.$ Clearly, any path
$\eta({\mathcal{L}})$ satisfies (\ref{des1}).

\begin{figure}[ptbh]
\includegraphics[width=0.48\textwidth]{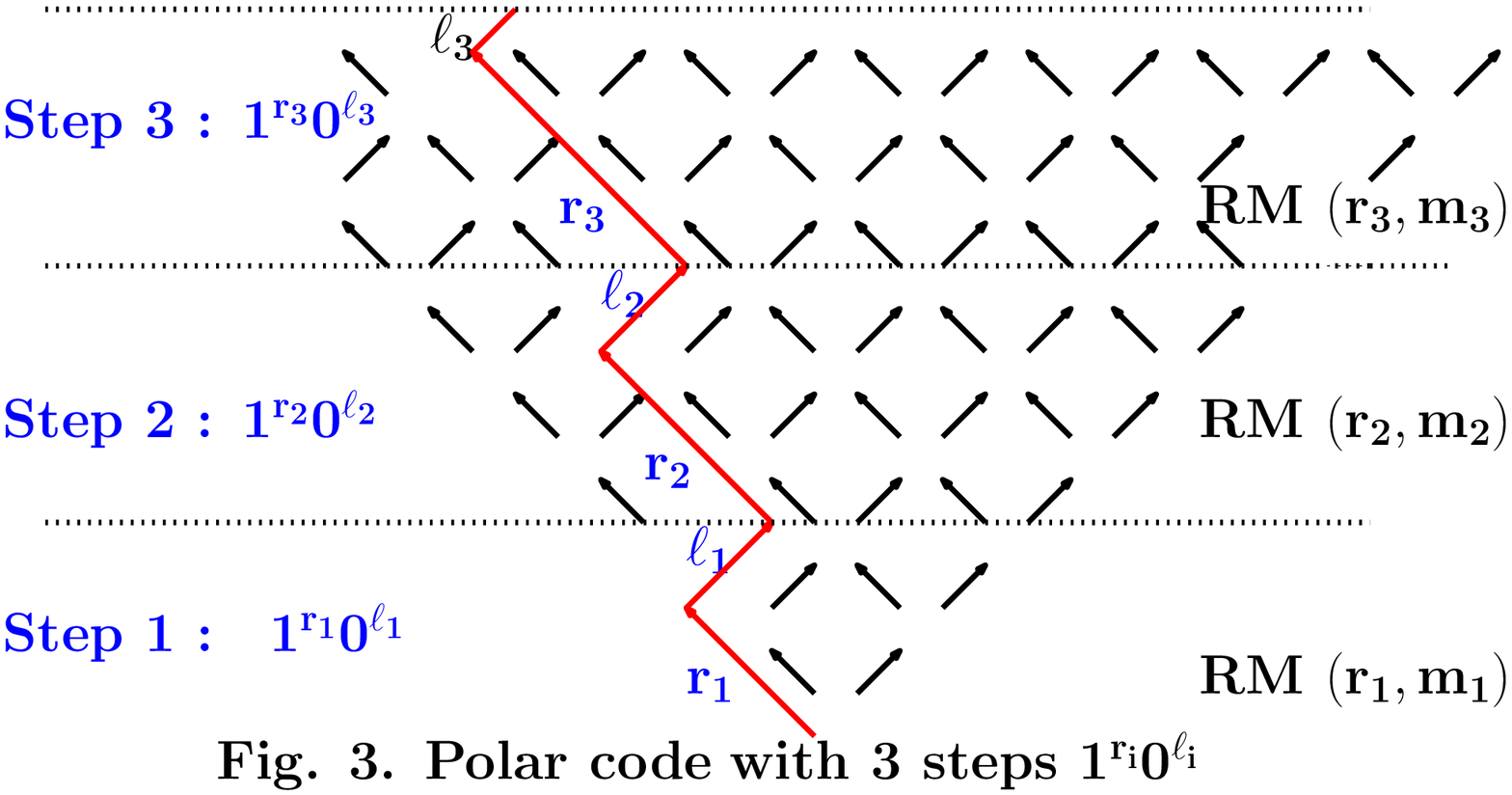}\end{figure}

\begin{lemma}
\label{lm:rate} Paths $\eta({\mathcal{L}})$ of (\ref{nu}) generate the direct
product%
\begin{equation}
\mathcal{R(}{\mathcal{L}})=\otimes_{i=1}^{s}\mathcal{R}(r_{i},m_{i})
\label{rate}%
\end{equation}
of $s$ RM codes $\mathcal{R}(r_{i},m_{i})$ of rates $R_{i}=k(r_{i}%
,m_{i})/2^{m_{i}}.$ Code $\mathcal{R(}{\mathcal{L}})$ has length $2^{m},$ code
rate $R_{{\mathcal{L}}}=\Pi_{i=1}^{s}\,\,R_{i},$ and construction complexity
of order $n\log n.$
\end{lemma}

\noindent\textit{Proof. }Each segment $\xi^{(i)}$ is a boundary for the
subpaths $\eta^{(i)}:\;w(\eta^{(i)})\leq{r}_{i}.$ These single-step subpaths
span the code $\mathcal{R}(r_{i},m_{i}),$ which is generated by monomials of
degree $r_{i}$ or less taken over variables $x_{M_{i}+1},...,x_{M_{i}+m_{i}},$
where $M_{i}=\sum_{j=1}^{i-1}m_{j},$ $M_{1}=0.$ Thus, $\mathcal{R(}%
{\mathcal{L}})$ is the direct product of codes $\mathcal{R}(r_{i},m_{i})$ and
has rate $R_{{\mathcal{L}}}.$ Each row $\eta$ of its generator matrix is a map
$\mathbf{x}^{\eta}:\mathbb{F}_{2}^{m}\rightarrow\mathbb{F}_{2}$ of the
monomial $\mathbf{x}^{\eta}$ defined by a path $\eta.$ Thus, conditions
(\ref{nu}) require $m$ operations to verify that $\eta\in\eta({\mathcal{L}})$
for any row $\eta.$ For $n$-row verification, the complexity is bounded by the
order $n\log n.$ \hfill
${\dimen0=1.5ex\advance\dimen0by-0.8pt\relax\mathop{\mkern0.5\thinmuskip
\vbox{\hrule width 1.5ex
\hbox to 1.5ex{\vrule
\hfill
\vrule height\dimen0 width 0pt
\vrule}\hrule width 1.5ex}\mkern0.5\thinmuskip}}$ \smallskip

Consider a sequence of channels BSC$(p)$ with $p\rightarrow0.$ Let $s=o\left(
\log m\right)  $ be some integer. We take a set of $2s$ numbers$\smallskip$%
\begin{equation}
{\mathcal{L=}}\left\{  {r}_{i}=2^{i-1}\log\left(  {}^{1}\!\!\left/
_{p}\right.  \right)  -c_{i},\;\;\ell_{i}=2^{i-1}\log\ts\log{}^{1}\!\!\left/
_{p}\right.  \right\}  \label{ell-1}%
\end{equation}
where $c_{1}=6,$ $c_{2}=3$ and $c_{i}=0$ for $i\geq3.$ We also assume that
${\mathcal{L}}$ is a set of integers. Then the path $\xi\left(  {\mathcal{L}%
}\right)  $ has the length$\smallskip$%
\begin{equation}
m=\ts\sum\nolimits_{i=1}^{s}m_{i}=(2^{s}-1)(\log{}^{1}\!\!\left/  _{p}\right.
+\log\log{}^{1}\!\!\left/  _{p}\right.  )-9 \label{dum6}%
\end{equation}
An equivalent setting arises if $p\sim m2^{-m/\left(  2^{s}-1\right)  }$ for
$m\rightarrow\infty.$ Note that the case $s=1$ gives a single code
$\mathcal{R}(r,m)$ with $m-r\sim\log m$ and $p\sim m2^{-m}.$ We first estimate
the redundancy $\rho_{{\mathcal{L}}}$ of a code $\mathcal{R(}{\mathcal{L}})$
with boundary (\ref{ell-1}).

\begin{lemma}
Codes $\mathcal{R(}{\mathcal{L}})$ with a boundary ${\mathcal{L}}$ of
(\ref{ell-1}) satisfy the redundancy bound (\ref{main}) for $p\to0$.\
\end{lemma}

\noindent\textit{Proof. }Let\textit{ }$\rho_{i}=1-R_{i}$ denote the redundancy
of code $\mathcal{R}(r_{i},m_{i})$ used in step $i$ of design (\ref{rate}).
Then
\[
\rho_{{\mathcal{L}}}=1-R_{{\mathcal{L}}}=1-\prod\nolimits_{1}^{s}(1-\rho
_{i})\leq\sum\nolimits_{1}^{s}\rho_{i}%
\]
\textit{ }Let\textit{ }$\tau=\log{}^{1}\!\!\left/  _{p}\right.  .$ First, note
that all codes $\mathcal{R}(r_{i},m_{i})$ have $\rho_{i}\rightarrow0$ as
$p\rightarrow0.$ Indeed, $\ell_{i}/m_{i}\leq\left(  \log\tau\right)
/\tau\rightarrow0$ and
\[
\rho_{i}\leq2^{-m_{i}}\left(  _{\,\ell_{i}}^{m_{i}}\right)  \leq
2^{-m_{i}[1-h(\ell_{i}/m_{i})]}%
\]
Second, note that each segment $\xi^{(i)}$ of boundary (\ref{ell-1}) has the
length $m_{i}\geq2m_{i-1}.$ Then $\rho_{i}\sim o(\rho_{i-1})$ and
$\rho_{{\mathcal{L}}}\sim\rho_{1}.$ Finally, we use the bounds
\begin{gather*}
\underset{}{h}(\ell_{i}/m_{i})=\ts\left(  \ell_{i}/m_{i}\right)  \log\left(
em_{i}/\ell_{i}\right)  +O\ts(\ell_{i}^{2}/m_{i}^{2})\smallskip\\
\rho_{1}\sim\left(  64p/\tau\right)  \ts\left(  e\tau/\log\tau\right)
^{\log\tau}<p\left(  \tau^{\log\tau}\right)  \smallskip
\end{gather*}
Thus, $\rho_{1}$ and $\rho_{{\mathcal{L}}}$ satisfy asymptotic bound
(\ref{main}). \hfill
${\dimen0=1.5ex\advance\dimen0by-0.8pt\relax\mathop{\mkern0.5\thinmuskip
\vbox{\hrule width 1.5ex
\hbox to 1.5ex{\vrule
\hfill
\vrule height\dimen0 width 0pt
\vrule}\hrule width 1.5ex}\mkern0.5\thinmuskip}}$ \smallskip

We now can prove Theorem 1 for construction (\ref{ell-1}). Here we use the
same approach that employed the boundary paths in \cite{dum1} and \cite{bd1};
however, we extend this approach to a multi-step boundary (\ref{ell-1})
instead of the single--step and double-step boundaries used before. \ We
proceed as follows. Consider any high-quality channel $W$, such as BSC$(p)$
with $p\rightarrow0$, and its two descendant channels $W^{(1)}$ and $W^{(0)}.$
Note that the degrading channel $W^{(1)}$ and the upgrading channel $W^{(0)}$
exhibit a vastly different behavior. In particular, let the original parameter
$Z(W)\sim\delta$ be close to 0 and the complementary parameter $\mathcal{G}%
(W)\in\lbrack1-\delta,(1-\delta^{2})^{1/2}]$ be close to 1. Then the channel
$W^{(0)}$ undergoes a sharp improvement over $W$ and yields an exponentially
declining parameter $Z(W^{(0)})\sim\delta^{2},$ according to (\ref{dum3}). By
contrast, the channel $W^{(1)}$ experiences a relatively small degradation and
yields $\mathcal{G}(W^{(1)})\in\lbrack1-2\delta,1-\delta^{2}].$ This allows us
to completely compensate the relatively long chains of degrading channels
$1^{{r}_{i}}$ with short chains $0^{\ell_{i}}$ of upgrading channels. In fact,
we will improve the overall performance in each step of the boundary
(\ref{ell-1}). It is this superiority of the chains $0^{\ell_{i}}$ that yields
small ratios $\ell_{i}/r_{i}$ in our design and leads to a nearly optimal
decline rate of redundancy $\rho_{{\mathcal{L}}}.$ The exact calculations are
given below. \

Consider two functions $f=f(n)$ and $r=r(n)$ that have the same sign. Then we
write $f\lesssim r$ or $f\gtrsim r$ if the asymptotic ratio $\lambda
=\lim_{n\rightarrow\infty}f/r$ is $\lambda\in(0,1)$ or $\lambda\geq1,$
respectively. We also write $f\succ r$ if $f>r^{c}$ for some $c>1.$ Finally,
consider inequalities \smallskip\
\begin{align}
-x-x^{2}  &  <\ln(1-x)<-x,\;\;x\in(0,^{1}\!\!\left/  _{2}\right.
)\label{log2}\\
1-x  &  <-\ln x,\;\;x\in(0,1)\nonumber
\end{align}
which are tight as $x\rightarrow0$ and $x\rightarrow1,$ respectively. Using
these inequalities, we can rewrite (\ref{dum1}) as
\begin{align}
\log Z(\xi)  &  <\underset{}{\tfrac{1}{2}}\log[-2\ln\mathcal{G}(\xi
)]\label{dum-z}\\
\ln\mathcal{G}(\xi)\;  &  >-Z(\xi)-Z^{2}(\xi) \label{dum-g}%
\end{align}
Below, we extensively use a recursion that employs inequalities (\ref{dum-z})
and (\ref{dum-g}). We will also see that $Z(\xi)\rightarrow0$ and
$\mathcal{G}(\xi)\rightarrow1$ for the selected path $\xi({\mathcal{L)}}$ of
(\ref{ell-1}). In this case, we can also replace (\ref{dum-z}) and
(\ref{dum-g}) with simpler inequalities $\log Z(\xi)\lesssim\tfrac{1}{2}%
\log\left[  -\ln\mathcal{G}(\xi)\right]  $ and $\ln\mathcal{G}(\xi
)\gtrsim-Z(\xi).$

\begin{lemma}
\label{th:path} Codes $\mathcal{R(}{\mathcal{L}})$ with a boundary
(\ref{ell-1}) achieve an output bit error rate $P_{\eta}\rightarrow0$ for each
path $\eta({\mathcal{L}})$ under SCD on a BSC$(p)$ with $p\rightarrow0.$
\end{lemma}

\noindent\textit{Proof. }Given the boundary $\xi\left(  {\mathcal{L}}\right)
,$ we will estimate the Bhattacharyya parameters%
\[
Z_{(i)}\equiv Z\left[  1^{{r}_{1}}0^{\ell_{1}}...1^{{r}_{i}}\right]
,\;Z^{(i)}\equiv Z\left[  1^{{r}_{1}}0^{\ell_{1}}...1^{{r}_{i}}0^{\ell_{i}%
}\right]
\]
obtained in processing of each step $i$. We also use similar notation
$\mathcal{G}_{(i)}$ and $\mathcal{G}^{(i)}$ for the offsets obtained in step
$i$. The original channel BSC$(p)$ gives parameter $\mathcal{G}=1-2p,$ where
$p\rightarrow0.$ For the first segment $1^{r_{1}},$ equality (\ref{dum2}) and
the upper bound (\ref{dum1}) give:$\smallskip$%
\begin{align}
\mathcal{G}_{(1)}  &  =(1-2p)^{1/\left(  64p\right)  }\sim e^{-1/32\smallskip
}\nonumber\\
Z_{(1)}  &  \lesssim\left(  1-e^{-1/16}\right)  ^{1/2}<2^{-2} \label{dum4}%
\end{align}
For the next segment $0^{\ell_{1}}$, equality (\ref{dum3}) gives%
\[
Z^{(1)}=\left[  Z_{(1)}\right]  {}^{2^{\ell_{1}}}<2^{-2\log{}^{1}\!\!\left/
_{p}\right.  }=p^{2}%
\]
Then $\mathcal{G}^{(1)}\geq1-Z^{(1)},$ according to (\ref{dum1}), and we
proceed with the segment $1^{{r}_{2}}0^{\ell_{2}}$ using (\ref{dum2}):
\begin{align*}
\mathcal{G}_{(2)}  &  \geq(1-p^{2})^{p^{-2}/8}\underset{}{\sim}e^{-1/8}\\
Z_{(2)}  &  \lesssim\left(  1-e^{-1/4}\right)  ^{_{1/2}}\underset{}{<}1/2\\
Z^{(2)}  &  =[Z_{(2)}]{}^{2^{\ell_{2}}}<2^{-\tau^{2}}\underset{}{=}p^{\tau}%
\end{align*}
Note that $2^{{r}_{i}}=p^{-2^{i-1}}$ and $2^{\ell_{i}}=\tau^{2^{i-1}}$ for
$i\geq3.$ Now we use inequalities (\ref{dum-z}) and (\ref{dum-g}) to prove
that parameters $Z^{(i)}$ rapidly decline$:$%
\begin{equation}
Z^{(i)}\leq p^{t_{i}},\;\;t_{i}=\tau^{2^{i}-i-1} \label{dum5}%
\end{equation}
Indeed, $Z^{(2)}\ $satisfies (\ref{dum5})$.$ We take $Z^{(i-1)}\leq
p^{t_{i-1}}$ and use induction on the $i$-th segment $1^{{r}_{i}}0^{\ell_{i}}%
$. Then inequalities (\ref{dum-z}) and (\ref{dum-g}) give$\smallskip$
\begin{align}
\ln\mathcal{G}_{(i)}\;\,  &  \geq-\underset{}{2}^{{r}_{i}}[p^{t_{i-1}%
}+p^{2t_{i-1}}]\gtrsim-2^{{r}_{i}}p^{t_{i-1}}\nonumber\\
\log Z_{(i)}  &  <\tfrac{1}{2}\log[-2\ln\mathcal{G}_{(i)}]\lesssim\underset
{}{\tfrac{1}{2}}\left(  t_{i-1}-{r}_{i}\right)  \log p \label{z0}%
\end{align}
$\smallskip$Note that ${r}_{i}=o(t_{i-1}).$ Thus, $\log Z_{(i)}\leq s_{i}\log
p,$ where
\[
s_{i}=t_{i-1}/\tau=\tau^{2^{i-1}-i-1}=o\left(  t_{i-1}\right)
\]
Then
\begin{equation}
\log Z^{(i)}=2^{\ell_{i}}\log Z_{(i)}\leq\tau^{2^{i-1}}s_{i}\log p=t_{i}\log p
\label{z}%
\end{equation}
This proves (\ref{dum5}) and gives $P_{\eta}\leq Z^{(s)}$ for each path $\eta
$. \hfill
${\dimen0=1.5ex\advance\dimen0by-0.8pt\relax\mathop{\mkern0.5\thinmuskip
\vbox{\hrule width 1.5ex
\hbox to 1.5ex{\vrule
\hfill
\vrule height\dimen0 width 0pt
\vrule}\hrule width 1.5ex}\mkern0.5\thinmuskip}}$ \smallskip

\textit{Discussion.} Inequalities (\ref{z0}) and (\ref{z}) show that the
initial chains $1^{r_{i}}$ and the subsequent chains $0^{\ell_{i}}$ affect
parameters $Z_{(i)}$ and $Z^{(i)}$ in a very different way. In particular,
(\ref{z0}) shows that any chain $1^{r_{i}}$ reduces the previous exponential
order $t_{i-1}=\log_{p}Z^{(i-1)}$ to $t_{i-1}/2-o(t_{i-1}).$ By contrast, the
stretch $0^{\ell_{i}}$ increases this order above $2^{\ell_{i}}(t_{i-1}%
/\tau).$ For this reason, good BSC$(p)$ with $p\rightarrow0$ may
overcompensate long chains $1^{r_{i}}$ of degrading channels with the much
shorter chains $0^{\ell_{i}}$ of upgrading channels. Note also that equalities
(\ref{dum2}) and (\ref{dum3}) are critical in our proof since they give exact
estimates $\mathcal{G}_{(i)}$ and $Z^{(i)}$ in all intermediate steps of the
segments $1^{r_{i}}{\ }${or} $0^{\ell_{i}},$ without any loss in performance.
To this end, note that inequalities (\ref{arik1}) and (\ref{dum1}) alone
cannot furnish Lemma \ref{th:path}. For example, inequalities (\ref{arik1})
replace estimate (\ref{dum4}) with a loose bound $Z_{(1)}\leq1-e^{-1/\left(
32\sqrt{p}\right)  }.$ This bound will require a much longer path $0^{\ell
_{1}}$ to achieve a low quantity $Z^{(2)}$, which in turn increases
redundancies $\rho_{1}$ and $\rho_{{\mathcal{L}}}$ above the bound
(\ref{dum1}) of the weakly optimal codes.

However, this particular construction fails to give the optimal redundancy
$\rho_{\text{opt}}\sim p\log{}^{1}\!\!\left/  _{p}\right.  $ or even reduce
$\rho_{{\mathcal{L}}}$ to the order of $cp\log{}{}^{1}\!\!\left/  _{p}\right.
$ for some constant $c>1.$ Nor is it known if other low-complexity algorithms
for polar or other codes can achieve $\rho_{\text{opt}}$ for $p\rightarrow0.$
Note also that the single-boundary set $\eta({\mathcal{L}})$ of Lemma
\ref{th:path} does not form an optimized polar code since many other paths
$\eta$ also have a vanishing output error rate. For example, any initial
segment $1^{{r}}$ of length $r<{r}_{1}$ gives rise to many paths $\eta
\notin\eta({\mathcal{L}}).$ To reduce redundancy $\rho_{{\mathcal{L}}}$, one
may consider a growing set $\left\{  \xi\right\}  $ of boundary paths $\xi$
and form an entire \textquotedblleft envelope" of the descendant paths
$\eta(\xi).$ Calculating the redundancy for this envelope-type boundary is
another open problem that may be related to the Young diagrams.$\vspace
{-0.06in}$

\section{\textbf{Low-rate} \textbf{codes with a stepped boundary}%
\label{sub:6}}

Consider a sequence of the BSCs$(p)$ with $p=(1-\epsilon)/2,$ where
$\epsilon\rightarrow0$ as length $n\rightarrow\infty.$ Below we study
capacity-achieving (CA) codes of rate $R\sim C$ for the case of a vanishing
capacity $C=1-h(p)\sim\epsilon^{2}/\ln4.$ It is proved in \cite{abbe-2015}
that RM codes $\mathcal{R}(r,\mu)$ are CA codes under ML-decoding if
$r=o(\mu).$ However, only codes $\mathcal{R}(1,\mu)$ of length $k=2^{\mu}$ or
their concatenations are known to be CA-codes of polynomial complexity. More
specifically, consider a BSC$(p_{\ast})$ with capacity $C\rightarrow0$ and
transition error probability
\begin{equation}
p_{\ast}=(1-\epsilon_{\ast})/2,\;\epsilon_{\ast}=\left(  C\ln4\right)  ^{1/2}
\label{sid}%
\end{equation}
According to \cite{sid-1992}, for any parameter $\theta\in(0,1),$ codes
$\mathcal{R}(1,\mu)$ of code rate $R=C(1-\theta)$ achieve on BSC$(p_{\ast})$
the output bit error rate $P_{\ast}\leq k^{-\theta}$ or less with complexity
$O(k\log k).$

To proceed with the low-rate codes, we need to substantially reduce the output
error rate of (\ref{dum5}). This is done in the following theorem, where we
reduce the error rate $P_{\eta}\leq Z^{(s)}$ at the expense of a slightly
higher redundancy $\rho_{{\mathcal{L}}}$. Consider a boundary$\smallskip$
\begin{equation}
{\mathcal{L}}_{c}{\mathcal{=}}\left\{  {r}_{i}=2^{i-1}\left(  \log{}%
^{1}\!\!\left/  _{p}\right.  \right)  -c_{i},\;\;\ell_{i}=c2^{i-1}\ts\log
{}^{1}\!\!\left/  _{p}\right.  \right\}  \label{ell2}%
\end{equation}
where $c_{1}=6,$ $c_{i}=0$ for $i\geq2,$ and $c\in(0,1)$ is a parameter. This
boundary has length
\begin{equation}
m=\sum\nolimits_{i=1}^{s}m_{i}=(c+1)\left(  2^{s}-1\right)  \left(  \log{}%
^{1}\!\!\left/  _{p}\right.  \right)  -6 \label{ell3}%
\end{equation}

\begin{lemma}
\label{th:path1} Codes $\mathcal{R(}{\mathcal{L}}_{c})$ with boundary
(\ref{ell2}) have redundancy $\rho_{{\mathcal{L}}}\rightarrow0$ as
$p\rightarrow0.$ These codes perform SCD with an output bit error rate
$P_{\eta},$ where for each path $\eta$,
\begin{equation}
\log P_{\eta}\lesssim-2^{2-s}p^{-c(2^{s}-1)} \label{z3}%
\end{equation}

\end{lemma}

\noindent\textit{Proof. }Note that $\ell_{i}/m_{i}=c/(c+1).$ For $i\geq2,$
let
\[
c_{1}\equiv h(\ell_{i}/m_{i})=h\left[  c/(c+1)\right]  <1.
\]
Then $\rho_{i}\leq2^{-m_{i}(1-c_{1})}=o(\rho_{i-1})$ for $p\rightarrow0,$ and
\[
\rho_{{\mathcal{L}}}\sim\rho_{1}\leq64p^{(1+c)(1-c_{1})}\rightarrow0
\]
Also, $2^{{r}_{i}}=p^{-2^{i-1}}$ and $2^{\ell_{i}}=p^{-c2^{i-1}}$ for
$i\geq2.$ Next, we estimate parameters $Z_{(i)}$ and $Z^{(i)}$ and follow the
proof of Lemma \ref{th:path}. Given the same length ${r}_{1},$ we again obtain
$Z_{(1)}<1/4$ of (\ref{dum4}). The next segment $0^{\ell_{1}}$ gives%
\[
Z^{(1)}=\left[  Z_{(1)}\right]  {}^{2^{\ell_{1}}}<2^{-2p^{-c}{}}%
\]
Then the segment $1^{{r}_{2}}0^{\ell_{2}}$ yields estimates
\begin{align}
\ln\mathcal{G}_{(2)}\;\,  &  \gtrsim-\underset{}{p}^{-2}Z^{(1)}\gtrsim
-p^{-2}2^{-2p^{-c}}\label{z2}\\
\log Z_{(2)}  &  \lesssim\underset{}{\tfrac{1}{2}}\log\left[  -2\ln
\mathcal{G}_{(2)}\right]  \lesssim-p^{-c}\nonumber\\
\log Z^{(2)}  &  \lesssim-\underset{}{2}^{\ell_{2}}p^{-c}\lesssim
-p^{-3c}\nonumber
\end{align}
Now we prove that parameters $Z^{(i)}$ rapidly decline$:$%
\begin{equation}
\log Z^{(i)}\lesssim-2^{2-i}p^{-c(2^{i}-1)} \label{z1}%
\end{equation}
Indeed, $Z^{(2)}\ $satisfies (\ref{z1})$.$ We take $Z^{(i-1)}$ of (\ref{z1})
and proceed with the $i$-th segment $1^{{r}_{i}}0^{\ell_{i}}$. We proceed
similarly to (\ref{z2}),
\begin{align*}
\ln\mathcal{G}_{(i)}\;\,  &  \gtrsim-\underset{}{2}^{{r}_{i}}Z^{(i-1)}\\
\log Z_{(i)}  &  <\tfrac{1}{2}\log\left[  -2\ln\mathcal{G}_{(i)}\right]
\underset{}{\lesssim}\tfrac{1}{2}{r}_{i}+\tfrac{1}{2}\log Z^{(i-1)}%
\end{align*}
$\smallskip$ Since ${r}_{i}=o(\log Z^{(i-1)}),$ we obtain :
\[
\log Z^{(i)}=2^{\ell_{i}}\log Z_{(i)}\lesssim2^{\ell_{i}-1}\log Z^{(i-1)}%
\]
which gives (\ref{z1}) and proves the theorem. \hfill
${\dimen0=1.5ex\advance\dimen0by-0.8pt\relax\mathop{\mkern0.5\thinmuskip
\vbox{\hrule width 1.5ex
\hbox to 1.5ex{\vrule
\hfill
\vrule height\dimen0 width 0pt
\vrule}\hrule width 1.5ex}\mkern0.5\thinmuskip}}\smallskip$

We will now combine codes $\mathcal{R}(1,\mu)$ with the high-rate polar codes
of Lemma \ref{th:path1} to obtain new CA codes.

Note that code $\mathcal{R}(1,\mu)$ is defined by a boundary path $\xi
^{(0)}=1^{1}0^{\mu-1}.$ We then combine $\xi^{(0)}$ with the boundary
${\mathcal{L}}_{c}$ of (\ref{ell2}) and obtain the extended boundary%
\begin{equation}
{\mathcal{L}}_{ext}=\left\{  {r}_{0}=1,\ell_{0}=\mu-1,\,{\mathcal{L}}%
_{c}\right\}  \label{ext1}%
\end{equation}
Lemma \ref{lm:rate} shows that ${\mathcal{L}}_{ext}$ generates the direct
product $\mathcal{R}_{ext}$ of $s+1$ RM codes $\mathcal{R}(r_{i},m_{i}).$
Thus, code $\mathcal{R}_{ext}$ has code rate $R$ and length $N,$ where
\begin{align*}
R  &  =R(1,\mu)R_{{\mathcal{L}}_{c}}\underset{}{\sim}(\mu+1)/2^{\mu}\\
N  &  =kn,\;\;k=2^{\mu},\;\;n=2^{m}%
\end{align*}
Codes $\mathcal{R}_{ext}$ also represent a simple concatenated construction,
which first uses $\mu+1$ arbitrary codewords of the code $\mathcal{R(}%
{\mathcal{L}}_{c})$ and forms an $\left(  \mu+1\right)  \times n$ matrix. Then
each column of this matrix is encoded into the code $\mathcal{R}(1,\mu).$ The
result is an $k\times n$ matrix, which represents a codeword formed by the
inner code of length $k$ and $s$ outer codes of length $n.$ Below we take
$\mu,m\rightarrow\infty.$ Below we take $p_{o}=2^{-\mu\theta}$ in
(\ref{ell2})$.$

\begin{theorem}
\label{th:low}Let codes $\mathcal{\mathcal{R}}_{ext}$ of code rate
$C(1-\theta)$ with an $s$-step boundary (\ref{ext1}) be used on a
BSC$(p_{\ast})$ of capacity $C\rightarrow0.$ For any $\theta\in(0,1),$ codes
$\mathcal{\mathcal{R}}_{ext}$ have decoding complexity $O(N\log N)$ in length
$N=nk$ and achieve a bit error probability $P_{\eta}$ such that
\begin{equation}
\log P_{\eta}\prec-2^{2-s}n^{c/(c+1)} \label{z4}%
\end{equation}

\end{theorem}

\noindent\textit{Proof.} Decoding of codes $\mathcal{R}_{ext}$ can be
expressed as SCD; below we also describe it as concatenated decoding of inner
codes. We take codes $\mathcal{R}(1,\mu)$ of rate $\epsilon^{2}/\ln4$ as
$\theta\rightarrow0$. Then $R_{ext}\sim\left(  1-\theta\right)  \epsilon
^{2}/\ln4$ as $m\rightarrow\infty.$ Given a received $2^{\mu}\times2^{m}$
matrix, we first perform ML decoding of each column of length $2^{\mu}$ into
the code $\mathcal{R}(1,\mu).$ The resulting $\left(  \mu+1\right)
\times2^{m}$ matrix contains errors with probability $p_{o}$ or less. Each row
is decoded into the code $\mathcal{R(}{\mathcal{L}})$ using SCD on a
BSC($p_{o}).$ Note that for any $\theta\in(0,1),$%
\begin{equation}
m=\left(  2^{s}-1\right)  \left(  c+1\right)  \mu\theta-c_{1} \label{m-1}%
\end{equation}
Then (\ref{z3}) gives the bit error rate (\ref{z4}):
\begin{equation}
\log P_{\eta}\lesssim-2^{2-s}2^{c\mu\theta(2^{s}-1)} \label{ext2}%
\end{equation}
Thus, codes $\mathcal{\mathcal{R}}_{ext}$ are CA codes. Inner and outer
decodings have the complexity $nk\log k$ and $\mu n\log n$ bounded by $N\log
N.$ \hfill
${\dimen0=1.5ex\advance\dimen0by-0.8pt\relax\mathop{\mkern0.5\thinmuskip
\vbox{\hrule width 1.5ex
\hbox to 1.5ex{\vrule
\hfill
\vrule height\dimen0 width 0pt
\vrule}\hrule width 1.5ex}\mkern0.5\thinmuskip}}$\smallskip

\textit{Discussion.} According to (\ref{ext2}), the order $\log P_{\eta}$
depends exponentially on the margin $\theta$ between the code rate $R$ and
channel capacity $C.$ Below, we compare the performance of codes
$\mathcal{R}(1,\mu)$ and $\mathcal{\mathcal{R}}_{ext}$ for the same code rate
$R\sim C(1-\theta),$ and define the minimum code length $k$ or $N$ that
enables a given output bit error $P.$ Here we consider an asymptotic case with
parameters $c\rightarrow1$ and $P\rightarrow0.$ For codes $\mathcal{R}%
(1,\mu),$ we have $k\sim P^{-1/\theta}$. For codes $\mathcal{\mathcal{R}%
}_{ext},$ we use notation $A=2^{s}\theta.$ Then parameters (\ref{m-1}) and
(\ref{ext2}) yield asymptotic approximations
\begin{equation}
n=2^{m}\asymp k^{2A},\;\log P\asymp-n^{1/2}\asymp-k^{A} \label{ext3}%
\end{equation}
(here $f\asymp r$ if $\log f\sim\log r).$ Recall that the outer codes
$\mathcal{R(}{\mathcal{L}}_{c})$ require a vanishing input error rate
$k^{-\theta}$, in which case $k=B^{1/\theta}$ for some $B\rightarrow\infty.$
Then $N=k^{2A+1}\asymp B^{1/\theta}(\log^{2}P).$ Thus, codes
$\mathcal{\mathcal{R}}_{ext}$ can improve the trade-off $k\sim P^{-1/\theta}$
of the inner codes $\mathcal{R}(1,\mu)$ only for the declining error rates
$P=o(1).$ We further note that this is the case for all other known
concatenated constructions. In particular, consider a classic concatenation
that uses the inner codes $\mathcal{R}(1,\mu)$ and the outer RS codes of the
same length $2^{\mu}$ and code rate $R\rightarrow1.$ It can be verified that
this construction requires the overall length $N_{1}\asymp\max\{\theta^{2}%
\log^{2}P,B^{2/\theta}\}$ given the same inner length $k=B^{1/\theta}.$ One
possible advantage of codes $\mathcal{\mathcal{R}}_{ext}$ over classic
concatenation is the extra parameter $s$ that allows the outer code length $n$
arbitrarily exceed the inner length $k$ in (\ref{m-1}). In particular, we have
inequality $N\lesssim N_{1}$ for both cases $B^{1/\theta}<\log^{2}P$ and
$B^{1/\theta}>\log^{2}P.$ Thus, construction of Theorem \ref{th:low} allows us
to shorten the length $N_{1}$ of the classical concatenated construction. More
generally, it is an important problem to find low-complexity codes of code
rate $R\rightarrow0$ that can achieve the vanishing error rates at the shorter
lengths of order $N\sim2^{c/\theta}$ for some $c\in(0,1).$

\section{Concluding remarks}

In this paper, we address explicit constructions of polar codes that are
nearly optimal for the extreme cases of a BSC($p)$ with $p\rightarrow0$ and
$p\rightarrow1/2.$ In case of $p\rightarrow0,$ we obtain weakly optimal codes
of rate $R\rightarrow1$, whose redundancy order $\log\rho$ declines at the
optimal rate (\ref{main}). For the low-rate codes, we obtain the optimal
decline of code rate $R\rightarrow0.$ These simple constructions are
completely defined by a single $s$-step boundary path $\xi\left(
{\mathcal{L}}\right)  $ that only depends on transition error probability $p.$
In turn, this boundary defines all other paths $\eta,$ which form other
sequences of upgrading-degrading channels included in code construction. An
important point is that the boundary ${\mathcal{L}}$ consists of the
consecutive chains of upgrading or degrading channels, with a growing length
of each segment. For this reason, these single-boundary codes can be
considered as direct products of $s$ Reed-Muller codes. One way to amplify
this design is to consider polar codes that include multiple overlapping
boundaries ${\mathcal{L}}_{1},...,{\mathcal{L}}_{k}$ and admit all descendant
paths $\eta$ that satisfy at least one boundary restriction. Another
interesting problem is to extend this design to other code rates and consider
the explicit constructions that admit the finite-length stretches of the
upgrading-degrading channels.

\noindent\textbf{Appendix.} \textit{Proof of Lemma \ref{lm:tau}. \ }To prove
Lemma \textit{\ref{lm:tau}}, we will assume that any channel $W_{\xi}$
satisfies \ the \textquotedblleft symmetry\textquotedblright\ condition
(\cite{rich}, p. 628). This condition (expressed in terms of log likelihoods
in \cite{rich}) implies that the likelihoods $h$ of transmitted symbols have
the probability density function (pdf) $p(x)\equiv p_{h}(x)$ such that
\begin{equation}
p(x)/p(x^{-1})=x,\quad\forall x\in(0,\infty). \label{pos0}%
\end{equation}
Condition (\ref{pos0}) can be used for many conventional channels; in
particular, for a BSC$(p)$ or an AWGN channel. It is also proven in
\cite{rich} that the \textquotedblleft symmetry\textquotedblright\ condition
is left intact by transformations (\ref{m2}). Namely, both rv $h_{j}^{(0)}$
and $h_{j}^{(1)}$ in (\ref{m2}) satisfy condition (\ref{pos0}) if so do rv
$h_{0,j}$ and $h_{1,j}$.

Next, we consider\textit{\ } an output $h(a)$ of the prefix path $a$. \ Let
$h_{1},h_{2},h_{3},$ and $h_{4}$ denote 4 independent ID rv$,$ which represent
4 different outputs $h(a)$ of the prefix $a.$ We need to calculate the outputs
$h_{01}\equiv h(a01)$ and $h_{10}\equiv h(a10)$ and prove that $\mathsf{E}%
h_{01}^{-\lambda}\leq\mathsf{E}h_{10}^{-\lambda}\;$if $\lambda\in\lbrack0,1].$
An equivalent formulation is to prove inequality $\mathsf{E}f_{01}^{\lambda
}\leq\mathsf{E}f_{10}^{\lambda}$ given \textit{ inverse} likelihoods
$f_{i}=h_{i}^{-1},$ $f_{01}=h_{01}^{-1}$ and $f_{10}=h_{10}^{-1}.$
Correspondingly, we consider the 4-dimensional space $\mathbb{R}_{+}^{4}$
\ formed by vectors $F=(f_{1},f_{2},f_{3},f_{4})$ with positive coordinates.
For extended subpaths $a01$ and $a10,$ recalculations (\ref{m2}) give the rv
outputs%
\[
f_{01}=\frac{f_{1}f_{2}+f_{3}f_{4}}{1+f_{1}f_{2}f_{3}f_{4}},
\]%
\[
f_{10}=\frac{(f_{1}+f_{2})(f_{3}+f_{4})}{(1+f_{1}f_{2})(1+f_{3}f_{4})}.
\]
Below we also consider another rv%
\[
u_{01}=\frac{(f_{1}+f_{2})(f_{3}+f_{4})}{2(1+f_{1}f_{2}f_{3}f_{4})}%
\]
and prove two inequalities
\begin{equation}
\mathsf{E}f_{01}^{\lambda}\leq\mathsf{E}u_{01}^{\lambda}\leq\mathsf{E}%
f_{10}^{\lambda},\;\lambda\in\lbrack0,1]. \label{tau0}%
\end{equation}
To prove the left inequality, note that $u_{01}=\left(  f_{01}^{\prime}%
+f_{01}^{\prime\prime}\right)  /2,$ where
\[
f_{01}^{\prime}=\frac{f_{1}f_{3}+f_{2}f_{4}}{1+f_{1}f_{2}f_{3}f_{4}}%
,\;f_{01}^{\prime\prime}=\frac{f_{1}f_{4}+f_{2}f_{3}}{1+f_{1}f_{2}f_{3}f_{4}%
}.
\]
The variables $f_{01}^{\prime}$ and $f_{01}^{\prime\prime}$ are obtained from
$f_{01}$ by replacements $f_{2}\Leftrightarrow f_{3}$ and $f_{2}%
\Leftrightarrow f_{4}$ respectively. Then independent and ID rv $f_{i}$ give
equalities
\[
\mathsf{E}f_{01}^{\lambda}=\mathsf{E}\left(  f_{01}^{\prime}\right)
^{\lambda}=\mathsf{E}\left(  f_{01}^{\prime\prime}\right)  ^{\lambda}%
\]
\ Since $x^{\lambda}$ is a concave function of any\ $x>0$ for $\lambda
\in\lbrack0,1],$
\begin{equation}
\ts\frac{\left(  f_{01}^{\prime}\right)  ^{\lambda}}{2}+\ts\frac{\left(
f_{01}^{\prime\prime}\right)  ^{\lambda}}{2}\leq\ts\left(  \frac
{f_{01}^{\prime}+f_{01}^{\prime\prime}}{2}\right)  ^{\lambda}=u_{01}^{\lambda}
\label{the2}%
\end{equation}
and $\mathsf{E}f_{01}^{\lambda}\leq\mathsf{E}u_{01}^{\lambda}.$

To compare the expectations $\mathsf{E}u_{01}^{\lambda}$ and $\mathsf{E}%
f_{10}^{\lambda},$ we combine \ each vector $F\equiv F_{0}\in\mathbb{R}%
_{+}^{4}$ with three other vectors (which may also coincide with $F):$
\begin{align*}
F_{1}  &  =(f_{1}^{-1},f_{2}^{-1},f_{3},f_{4}),\,\;F_{2}=(f_{1}^{-1}%
,f_{2}^{-1},f_{3}^{-1},f_{4}^{-1}),\,\\
F_{3}  &  =(f_{1},f_{2},f_{3}^{-1},f_{4}^{-1})
\end{align*}
We also consider the \textit{orbit} $\mathbf{T}=\{F_{0},F_{1},F_{2},F_{3}\}$
of vector $F_{0}\in\mathbb{R}_{+}^{4}$. Clearly, the whole space
$\mathbb{R}_{+}^{4}$ is now partitioned into non-intersecting orbits
$\mathbf{T}$. Below we use notation
\[
\alpha=f_{1}f_{2},\quad\beta=f_{3}f_{4},\;A=(f_{1}+f_{2})(f_{3}+f_{4}).
\]
It can be readily verified that the rv $f_{10}(\mathbf{T})$ does not change on
the orbit $\mathbf{T}$:
\begin{equation}
f_{10}(F_{i})=\frac{A}{\left(  1+\alpha\right)  (1+\beta)},\;i=0,...,3,
\label{tau1}%
\end{equation}
while $u_{01}(\mathbf{T})$ takes two values
\begin{align*}
u_{01}(F_{0})  &  =u_{01}(F_{2})=\frac{A}{2(1+\alpha\beta)}\\
u_{01}(F_{1})  &  =u_{01}(F_{3})=\frac{A}{2(\alpha+\beta)}%
\end{align*}
Let $p=p(F_{0})$ denote the pdf of the 4-dimensional rv $F_{0}\in
\mathbb{R}_{+}^{4},$ which consists of inverse likelihoods. According to
(\ref{pos0}), the pdfs of other orbit points are
\begin{equation}
p\left(  F_{2}\right)  =\alpha\beta p,\;p\left(  F_{1}\right)  =\alpha
p,\;p\left(  F_{3}\right)  =\beta p \label{pos1}%
\end{equation}
Then simple recalculations using equalities (\ref{tau1}) and (\ref{pos1})
give
\begin{align*}
\mathsf{E}f_{10}^{\lambda}(\mathbf{T})  &  =pA^{\lambda}\left[  (1+\alpha
)(1+\beta)\right]  ^{1-\lambda},\\
\mathsf{E}u_{01}^{\lambda}(\mathbf{T})  &  =pA^{\lambda}2^{-\lambda}\left[
\left(  1+\alpha\beta\right)  ^{1-\lambda}+\left(  \alpha+\beta\right)
^{1-\lambda}\right]
\end{align*}
Since $x^{1-\lambda}$ is a concave function$,$ we have inequality
\begin{equation}
\ts\frac{\left(  1+\alpha\beta\right)  ^{1-\lambda}}{2}+\ts\frac{\left(
\alpha+\beta\right)  ^{1-\lambda}}{2}\leq\ts\left[  \frac{(1+\alpha)(1+\beta
)}{2}\right]  ^{1-\lambda} \label{the6}%
\end{equation}
which proves the right inequality in (\ref{tau0}). The second case with
$\lambda\in\lbrack1,\infty)$ is studied similarly. Now both $x^{\lambda}$ and
$x^{1-\lambda}$ are a convex functions of \ $x>0$. Then inequalities
(\ref{the2}) and \ (\ref{the6}) change their sign and we have inequality
(\ref{ch3}). \hfill\ $\square$\smallskip$\frac{{}}{{}}$

\end{document}